\documentclass[12pt]{article}
\pdfoutput = 1
\begin{document}
%\date{Today}
\title
{Generalized uncertainty principle and black hole thermodynamics}

\author{
{\bf {\normalsize Sunandan Gangopadhyay}$^{a,b,c}
$\thanks{sunandan.gangopadhyay@gmail.com, sunandan@iucaa.ernet.in}},
{\bf {\normalsize Abhijit Dutta}
$^{d}$\thanks{dutta.abhijit87@gmail.com}},
{\bf {\normalsize Anirban Saha }
$^{b,c}$\thanks{anirban@iucaa.ernet.in}}\\
$^{a}$ {\normalsize National Institute for Theoretical Physics, Stellenbosch University, South Africa}\\
$^{b}${\normalsize Department of Physics, West Bengal State University, Barasat, Kolkata 700126, India}\\
$^{c}${\normalsize Visiting Associate in Inter University Centre for Astronomy $\&$ Astrophysics,}\\
{\normalsize Pune, India}\\
$^{d}${\normalsize Department of Physics, Adamas Institute of Technology, Barasat, Kolkata 700126, India}\\[0.3cm]
}
\date{}

\maketitle

\begin{abstract}
{\noindent We study the Schwarzschild and Reissner-Nordstr\"{o}m  black hole thermodynamics using the
simplest form of the generalized uncertainty principle (GUP) proposed in the literature. 
The expressions for the mass-temperature relation, heat capacity and entropy are obtained in both cases
from which the critical and remnant masses are computed.
Our results are exact and reveal that these masses are identical and larger than the so called singular mass for which
the thermodynamics quantities become ill-defined. The expression for the entropy reveals the well known area theorem
in terms of the horizon area in both cases upto leading order corrections from GUP. 
The area theorem  written in terms of a new variable which can be interpreted as the reduced horizon area arises  only when the computation is carried out to the next higher order
correction from GUP.

}

\end{abstract}

%\pacs{11.10.Nx, 03.65.Ta, 11.10.Ef, 04.30.Nk, 42.50.Dv}
\maketitle

%%%%%%%%%%%%%%%%%%%%%%%%%%%%%%%%%%%%%%%%%%%%%%%
%%%%%%%%%%%%%%%%%%%%%%%%%%%%%%%%%%%%%%%%%%%%%%%

\noindent Various theories of quantum gravity have suggested the need of an observer-independent
minimum length scale. A minimal length expected to be close or
equal to the Planck length occurs in string theory \cite{str}, noncommutative geometry \cite{nc},
to name a few. One of the manifestations of the inclusion of a minimal length in these theories
is the generalized uncertainty principle (GUP). This idea, first proposed in \cite{mead} has recently
been considered seriously to study black hole thermodynamics \cite{hawk1}-\cite{bek} and its quantum corrected entropy \cite{adler}-\cite{rb}, compute quantum gravity corrections in quantum systems (such as particle in a box, Landau levels,
simple harmonic oscillator, etc.)\cite{das1}-\cite{das6},
compute Planck scale  corrections to the phenomena of superconductivity and quantum Hall effect \cite{saurya}
and to understand its consequences in cosmology \cite{das7}, \cite{das8}.

\noindent In this paper, we shall study the thermodynamic properties of the Schwarzschild and Reissner-Nordstr\"{o}m (RN)
black holes using the simplest form of the GUP proposed in the literature \cite{adler}\footnote{The implications
of the GUP in quantum gravity phenomenology has been investigated recently in \cite{das1}, \cite{das2}.}
\begin{eqnarray}
\delta x\delta p\geq\frac{\hbar}{2}\left\{1+ \frac{\beta^2 l_p^2}{\hbar^2}(\delta p)^2\right\}
\label{gup}
\end{eqnarray}
where $l_p$ is the Planck length ($\sim 10^{-35}m$) and $\beta$ is a dimensionless constant.
We compute the critical and the singular
masses for these black holes below which the thermodynamic quantities become ill-defined. We then find the remnant
mass at which the radiation process stops. All our results are analytically exact and hence differ from \cite{rb}
where approximations have been made in the computation of the critical and remnant masses. 
We finally compute the entropy and obtain the well known area theorem.
For the Schwarzschild black hole, we go one step further to compute the entropy to the next leading order correction
from the GUP and obtain the area theorem in terms of a new variable which can be interpreted as the reduced horizon area.
The reduced horizon area leads to a singularity in the entropy which is avoided as the remnant mass is larger than the
mass for which the entropy becomes singular. It is to be noted that the reduced horizon area 
does not come in the picture if we keep our computation to the leading order correction from GUP.

\noindent To begin with, let us consider a Schwarzschild black hole of mass $M$. For any quantum particle (massless) 
near the horizon of a black hole, the momentum uncertainty characterizing its temperature can be written as \cite{adler}
\begin{eqnarray}
T=\frac{(\delta p) c}{k_B}
\label{e1}
\end{eqnarray}
where $c$ is the speed of light and $k_B$ is the Boltzmann constant.
For thermodynamic equilibrium, the temperature of the particle 
will be identical to the temperature of the black hole itself.
To relate this temperature with the mass of the black hole, we need to recast the GUP (\ref{gup}) 
in terms of $T$ and $M$ which in turn implies that the GUP (\ref{gup}) has to be saturated
\begin{eqnarray}
\delta x\delta p=\frac{\hbar}{2}\left\{1+ \frac{\beta^2 l_p^2}{\hbar^2}(\delta p)^2\right\}~.
\label{e3}
\end{eqnarray}
Near the horizon of the Schwarzschild black hole, the position uncertainty of a particle will be of the order of the Schwarchild radius
of the black hole \cite{adler},\cite{medved}
\begin{eqnarray}
\delta x&=&\epsilon r_{s}~;~
r_{s}=\frac{2GM}{c^2}
\label{e2}
\end{eqnarray}
where $\epsilon$ is a scale factor, $r_s$ is the Schwarzschild radius and $G$ is the Newton's universal gravitational constant.

\noindent  Substituting the value of  $\delta p$ and $\delta x$ from eqs.(\ref{e1}) and (\ref{e2}),
the GUP (\ref{e3}) can be rewritten as
\begin{eqnarray}
M&=&\frac{M_{p}^2 c^2}{4\epsilon k_{B}T}\left\{1+ \frac{\beta^2 l_p^2}{c^2 \hbar^2}(k_{B}T)^2\right\}\nonumber\\
&=&\frac{M_{p}^2 c^2}{4\epsilon}\left\{\frac{1}{k_B T}+\beta^2\frac{k_{B}T}{(M_p c^2)^2}\right\}
\label{e4}
\end{eqnarray}
where we have used the relations  $\frac{c\hbar}{l_p}=M_{p}c^{2}$ and $M_{p}=\frac{c^2 l_p}{G}$,  $M_{p}$ 
being the Planck mass.

\noindent In the absence of correction due to quantum gravity, eq.(\ref{e4}) reduces to
\begin{eqnarray}
M=\frac{M_{p}^2 c^2}{4\epsilon k_{B}T}~.
\label{e5}
\end{eqnarray}
Thus, comparing with the semi-classical Hawking temperature \cite{hawk1}, \cite{hawk2} $T=\frac{M_{p}^{2}c^2}{8 \pi M k_{B}}$, yields the value of $\epsilon=2\pi$.

\noindent This finally fixes  the form of the mass-temperature relation (\ref{e4}) to be
\begin{eqnarray}
M=\frac{M_{p}^2 c^2}{8\pi}\left\{\frac{1}{k_B T}+\beta^2\frac{k_{B}T}{(M_p c^2)^2}\right\}~.
\label{e6}
\end{eqnarray}
Now the heat capacity of the black hole can be defined as
\begin{eqnarray}
C=c^2\frac{dM}{dT}~.
\label{e7}
\end{eqnarray}
Using eq.(\ref{e6}), we get
\begin{eqnarray}
C=\frac{k_B}{8\pi}\left\{-\left(\frac{M_p c^2}{k_B T}\right)^{2} + \beta^2\right\}~.
\label{e8}
\end{eqnarray}
From eq.(\ref{e8}), it is obvious that for temperatures for which $k_B T<<M_p c^2$, 
the heat capacity is negative since the first term which comes with a negative sign dominates over the second term.
The radiation process leads to a decrease in the mass of the black hole which in turn (from eq.(\ref{e5}))
leads to an increase in the temperature. Now as the temperature increases, the heat capacity also increases and
it is easy to see from eq.(\ref{e8}) that there will be a point at which the heat capacity vanishes. We
consider the corresponding temperature to be the maximum temperature attainable by a black hole during evaporation. 
There will be no further change of black hole mass with its temperature.
The radiation process stops thereafter with a finite remnant mass with a finite temperature.

\noindent One can also determine the black hole entropy from the first law of black hole thermodynamics given by
\begin{eqnarray}
S=\int c^2 \frac{dM}{T}=\int C\frac{dT}{T}
\label{e8a}
\end{eqnarray}
where the second equality follows from eq.(\ref{e7}). Using eq.(\ref{e8}) and performing the above integration leads to
\begin{eqnarray}
S=\frac{k_B}{8 \pi}\left\{\frac{1}{2}\left(\frac{M_p c^2}{k_B T}\right)^2 
+ \beta^2 \ln\left(\frac{k_B T}{M_p c^2}\right)\right\}~.
\label{e8b}
\end{eqnarray}
Now to express the heat capacity and the entropy in terms of the mass, we need to obtain an expression for
the temperature $T$ in terms of the mass $M$. In order to proceed, we first define for 
notational convenience the following relations
\begin{eqnarray}
M'=\frac{8\pi M}{M_p}~;~T'=\frac{k_B T}{M_p c^2}~. 
\label{e8c}
\end{eqnarray}
Using these definitions, eq.(\ref{e6}) can be recast as
\begin{eqnarray}
M'=\frac{1}{T'}+\beta^2 T'  
\label{e8d}
\end{eqnarray}
which leads to the following quadratic equation for $T'$
\begin{eqnarray}
\beta^2 T'^2 -M' T' +1=0.
\label{quadeqn}
\end{eqnarray}
Solving this equation gives the following expression for $T' (T)$ in terms of $M' (M)$   
\begin{eqnarray}
T'=\frac{1}{2\beta^2}\left\{M'-\sqrt{M'^2-4\beta^2}\right\}~.
\label{sol}
\end{eqnarray}
The negative sign has been taken before the square root so that the solution reproduces eq.(\ref{e5}) in the $\beta\rightarrow0$ limit.
The above relation immediately gives a critical mass below which the temperature becomes a complex
quantity
\begin{eqnarray}
M_{cr}=\frac{\beta}{4\pi}M_p ~.
\label{sol_crit}
\end{eqnarray}
We would like to mention that our expression for the critical mass differs from that in \cite{rb}. The reason for this
discrepancy is due to the fact that in writing the temperature in terms of the mass, eq.(\ref{e8d}) has been squarred and the
term of the $\mathcal{O}(\beta^4)$ has been dropped leading to a different expression for the critical mass.

\noindent Squarring eq.(\ref{sol}) and substituting in eq.(\ref{e8}) leads to
\begin{eqnarray}
C=\frac{k_B}{8\pi}\left\{-\frac{2\beta^4}{M'^2 -2\beta^2 -M'\sqrt{M'^2 - 4\beta^2}} + \beta^2\right\}~.
\label{heatcap}
\end{eqnarray}
This relation gives the variation of the heat capacity with mass. To obtain the remnant mass at which the
radiation process terminates, we set $C=0$ which yields the following equation
\begin{eqnarray}
-\frac{2\beta^4}{M'^2 -2\beta^2 -M'\sqrt{M'^2 - 4\beta^2}} + \beta^2 =0~.
\label{hc}
\end{eqnarray}
Solving this equation gives
\begin{eqnarray}
M_{rem}=\frac{\beta}{4\pi}M_p ~.
\label{rem}
\end{eqnarray}
Once again the expression for the remnant mass differs from that in \cite{rb}.
Interestingly, we find that the remnant and the critical masses are identical. This is in contrast to
the results found earlier where they were different \cite{rb}. The reason for this discrepancy is due to the fact that
our results are exact in contrast to \cite{rb} where approximations (mentioned earlier) have been carried out in writing the temperature in terms of the mass from the mass-temperature relationship.

\noindent We now move on to express the entopy (\ref{e8b}) in terms of the mass.
Substituting eq.(\ref{sol}) in eq.(\ref{e8b}) and carrying out a binomial expansion keeping 
terms upto  $\mathcal{O}(\beta^2)$ leads to
\begin{eqnarray}
\frac{S}{k_B} &=&\frac{4\pi M^2}{M_{p}^{2}} -\frac{\beta^2}{8\pi}\ln\left(\frac{8\pi M}{M_p}\right)
-\frac{\beta^2}{8\pi}\nonumber\\
&=&\frac{S_{BH}}{k_B} - \frac{\beta^2}{16\pi}\ln\left(\frac{S_{BH}}{k_B}\right)
-\frac{\beta^2}{16\pi}\ln(16\pi)-\frac{\beta^2}{8\pi}
\label{e12}
\end{eqnarray}
where  $\frac{S_{BH}}{k_B} =\frac{4 \pi M^2}{M_p^2}$ is the semi-classical Bekenstein-Hawking entropy
for the Schwarzschild black hole.

\noindent  In terms of the area of the horizon 
$A=4\pi r_{s}^2 =16\pi \frac{G^2 M^2}{c^4} = 4 l_p^2 \frac{S_{BH}}{k_B}$,
eq.(\ref{e12}) can be written as
\begin{eqnarray}
\frac{S}{k_B} = \frac{A}{4l_p^2} -\frac{\beta^2}{16\pi}\ln\left(\frac{A}{4l_{p}^{2}}\right)
-\frac{\beta^2}{16\pi}\ln(16\pi)-\frac{\beta^2}{8\pi}
\label{e13}
\end{eqnarray}
which in the limit $\beta\rightarrow0$ is the famous area theorem. Interestingly, carrying out the calculation upto $\mathcal{O}(\beta^4)$ yields
\begin{eqnarray}
\frac{S}{k_B} =\frac{4\pi M^2}{M_{p}^{2}} -\frac{\beta^2}{8\pi}\ln\left(\frac{8\pi M}{M_p}\right)
-\frac{\beta^2}{8\pi}+\frac{\beta^2}{8\pi}\ln\left(1+\frac{\beta^2 M_{p}^2}{64\pi^2 M^2}\right)
+\frac{3}{16\pi}\frac{\beta^4 M_{p}^2}{64\pi^2 M^2}~.
\label{e12a}
\end{eqnarray}
Introducing a new variable $A'$  defined as
\begin{eqnarray}
A' &=& 16\pi\frac{G^2 M^2}{c^4}-\frac{\beta^2}{2\pi}\frac{G^2 M_{p}^2}{c^4}\nonumber\\
&=&A - \frac{\beta^2}{2 \pi}l_{p}^2
\label{e14}
\end{eqnarray}
one can rewrite eq.(\ref{e12a}) as
\begin{eqnarray}
\frac{S}{k_B} = \frac{A'}{4 l_p^2} - \frac{\beta^2}{16 \pi} \ln\left( \frac{A'}{4 l_p^2}\right)
+\frac{3\beta^4}{64\pi^2}\frac{l_{p}^2}{A'}-\frac{\beta^2}{16\pi}\ln(16\pi)~.
\label{e15}
\end{eqnarray}
This is the area theorem in terms of the variable $A'$ which can be interpreted as the reduced horizon area.
Note that the entropy has a singularity at  zero reduced horizon area which corresponds to a singular mass given by        
\begin{eqnarray}
M_{sing}=\frac{\beta}{\sqrt{32}\pi}M_{p}~.
\label{singmass}
\end{eqnarray}         
However, the remnant mass (\ref{rem})  is larger than the above mass and 
therefore the reduced horizon area is always positive which in turn implies that the 
black hole never approaches the singularity. This concludes our study of the effect of the GUP in the
thermodynamics of the Schwarzschild black hole.

\noindent We now consider the Reissner-Nordstr\"{o}m black hole of mass $M$ and charge $Q$.
In this case, near the horizon of the black hole, the position uncertainty of a particle will be of the order of the 
RN radius  of the black hole
\begin{eqnarray}
\delta x&=&\epsilon r_h\nonumber\\
r_h &=& \frac{Gr_0}{c^2}\nonumber\\
r_0 &=& M+\sqrt{M^2 -Q^2}
\label{rn1}
\end{eqnarray}
where $r_h$ is the radius of the horizon of the RN black hole. Substituting the value of 
$\delta p$ and $\delta x$ from eqs.(\ref{e1}) and (\ref{rn1}),
the GUP (\ref{e3}) can be rewritten as
\begin{eqnarray}
r_0 =\frac{M_{p}^2 c^2}{2\epsilon}\left\{\frac{1}{k_B T}+\beta^2\frac{k_{B}T}{(M_p c^2)^2}\right\}~.
\label{rn2}
\end{eqnarray}
This equation gives the relation between the mass, charge and temperature of the RN black hole.
Once again, in the absence of correction due to quantum gravity eq.(\ref{rn2}) reduces to
\begin{eqnarray}
r_0=\frac{M_{p}^2 c^2}{2\epsilon k_{B}T}~.
\label{rn3}
\end{eqnarray}
Comparing with the semi-classical Hawking temperature  
$T=\frac{M_{p}^{2}c^2 (Mr_0 -Q^2)}{2\pi k_{B}r_{0}^{3}}$, yields the
value of $\epsilon$ to be
\begin{eqnarray}
\epsilon=\frac{\pi r_{0}^2}{(Mr_0 -Q^2)}~.
\label{rn3a}
\end{eqnarray}
This finally fixes  the form of the mass-charge-temperature relation (\ref{rn2}) to be
\begin{eqnarray}
\frac{r_{0}^2}{(r_0 -M)}=\frac{M_{p}^2 c^2}{2\pi}\left\{\frac{1}{k_B T}+\beta^2\frac{k_{B}T}{(M_p c^2)^2}\right\}
\label{rn4}
\end{eqnarray}
where we have made use of the identity
\begin{eqnarray}
\frac{r_{0}}{(Mr_0 -Q^2)}=\frac{1}{(r_0 -M)}~.
\label{rn4a}
\end{eqnarray}
The heat capacity of the black hole can now be computed using relation (\ref{e7}) and eq.(\ref{rn4}) and gives
\begin{eqnarray}
C=\frac{k_B (r_0 -M)^3}{2\pi r_{0}^2(2r_0 -3M)}\left\{-\left(\frac{M_p c^2}{k_B T}\right)^2 +\beta^2\right\}~.
\label{rn5}
\end{eqnarray}
To express the heat capacity in terms of the mass, once again we make use of the relation (\ref{e8c}) to recast
eq.(\ref{rn4}) in the form
\begin{eqnarray}
\beta^{2} T'^{2} -\frac{g(r_0)}{M_p}T' +1=0\\
g(r_0)=\frac{2\pi r_{0}^2}{(r_0 -M)}\nonumber~.
\label{rn6}
\end{eqnarray}
Solving this equation gives the following expression for $T' (T)$ in terms of the mass and the charge of the RN black hole 
\begin{eqnarray}
T'=\frac{g(r_0)}{2\beta^2 M_p}\left\{1-\sqrt{1-\frac{4\beta^2 M_{p}^2}{g^{2}(r_0)}}\right\}~.
\label{rn7}
\end{eqnarray}
The negative sign has been taken again before the square root so that the solution reproduces eq.(\ref{sol}) in the $Q\rightarrow0$ limit.
The above relation immediately gives the following condition for the temperature $T'(T)$ to be real
\begin{eqnarray}
1-\frac{4\beta^2 M_{p}^2}{g^{2}(r_0)}\geq0~.
\label{rn8}
\end{eqnarray}
Taking the equality sign, we finally obtain the following cubic equation for the critical mass below which the temperature 
becomes a complex quantity
\begin{eqnarray}
4aM_{cr}^3 -a^2 M_{cr}^2 -4aQ^2 M_{cr} +Q^4 +a^2 Q^2=0~;~a=\frac{\beta M_p}{\pi}~.
\label{rn9}
\end{eqnarray}
Solving the above equation, we get
\begin{eqnarray}
M_{cr}=\frac{\beta M_p}{12\pi}\left\{1+\frac{(\beta^2 M_{p}^{2} +48\pi^2 Q^2)}
{A^{1/3}}+\frac{A^{1/3}}{\beta^2 M_{p}^{2}}\right\}~.
\label{rn10}
\end{eqnarray}
where 
\begin{eqnarray}
A=\beta^6 M_{p}^6 -144 \beta^4 M_{p}^{4} \pi^2 Q^2 -216\beta^2 M_{p}^2 \pi^4 Q^4 
+12\sqrt{3}\beta^2 M_{p}^2(\beta^2 M_{p}^{2}\pi Q -2\pi^3 Q^3)(27\pi^2 Q^2 -\beta^2 M_{p}^2)^{1/2}.
\label{rn11}
\end{eqnarray}
The above expression for the critical mass reduces to the critical mass for the Schwarzschild black hole (\ref{sol_crit})
in the $Q\rightarrow0$ limit.

\noindent Squarring eq.(\ref{rn7}) and substituting in eq.(\ref{rn5}) leads to
\begin{eqnarray}
C=\frac{k_B (r_0 -M)^3}{2\pi r_{0}^2(2r_0 -3M)}\left\{-\frac{4\beta^4 M_{p}^2}{g^{2}(r_0)}\frac{1}{\left\{1-\sqrt{1-\frac{4\beta^2 M_{p}^2}{g^{2}(r_0)}}\right\}^2}+\beta^2\right\}~.
\label{rn12}
\end{eqnarray}
This relation gives the variation of the heat capacity with mass for the RN black hole. To obtain the remnant mass at which the
radiation process terminates, we set $C=0$ which yields the same cubic equation 
for the remnant as that for the critical mass (\ref{rn9}).
Hence, the remnant mass is once again equal to the critical mass (for the RN black hole) 
similar to the Schwarzschild black hole.

\noindent Finally, we proceed to compute the entropy of the RN black hole. To do that,
we substitute the expressions for the temperature $T'(T)$ (eq.(\ref{rn7}))  
and heat capacity  (eq.(\ref{rn12})) in eq.(\ref{e8a}) 
which yields upto $\mathcal{O}(\beta^2)$
\begin{eqnarray}
\frac{S}{k_{B}}&=&2\beta^2 \int \frac{dM}{g(r_0)\left\{1-\sqrt{1-\frac{4\beta^2 M_{p}^2}{g^{2}(r_0)}}\right\}}\nonumber\\
&=&\frac{\pi r_{h}^2}{l_{p}^2}-\frac{\beta^2}{16\pi}\ln\left(\frac{\pi r_{h}^2}{l_{p}^2}\right)
-\frac{\beta^2 Q^2}{8M_p \left(\frac{\pi r_{h}^2}{l_{p}^2}\right)}\left\{1-\frac{\pi Q^2}{4M_{p}\left(\frac{\pi r_{h}^2}{l_{p}^2}\right)}\right\}\nonumber\\
&=&\frac{S_{BH}}{k_{B}}-\frac{\beta^2}{16\pi}\ln\left(\frac{S_{BH}}{k_{B}}\right)
-\frac{\beta^2 Q^2}{8M_p \left(\frac{S_{BH}}{k_{B}}\right)}\left\{1-\frac{\pi Q^2}{4M_{p}\left(\frac{S_{BH}}{k_{B}}\right)}\right\}
\label{rn_entr}
\end{eqnarray}
where  $\frac{S_{BH}}{k_B} =\frac{\pi r_{h}^2}{l_{p}^2}$ is the semi-classical 
Bekenstein-Hawking entropy for the RN black hole.
\noindent  In terms of the area of the horizon $A =4\pi r_{h}^2 =4l_{p}^2 \frac{S_{BH}}{k_B}$,
the above equation can be written as
\begin{eqnarray}
\frac{S}{k_B} = \frac{A}{4l_p^2} -\frac{\beta^2}{16\pi}\ln\left(\frac{A}{4l_{p}^{2}}\right)
-\frac{\beta^2 Q^2}{8M_p \left(\frac{A}{4l_{p}^2}\right)}\left\{1-\frac{\pi Q^2}{4M_{p}\left(\frac{A}{4l_{p}^2}\right)}\right\}
\label{entr_area}
\end{eqnarray}
which is the area theorem for the RN black hole.

\noindent To conclude, we now compare our results with earlier findings to put our results in a proper perspective.
The existence of critical and remnant masses were also found earlier in \cite{rb} for the Schwarzschild black hole. In
this paper, we present these results for both the Schwarzschild and the RN black holes. Our
expressions differ from the earlier findings and reveal that these masses are identical in both cases.
The reason for this difference is that the computations carried out in \cite{rb} were approximate in contrast
to our paper where the computations are exact. The expression for the entropy exhibits the well known area theorem
in terms of the horizon area in both cases upto leading order corrections from GUP. 
The computation carried out to the next higher order
correction from GUP (in case of the Schwarzschild black hole) shows that the area theorem can be written
in terms of a new variable which can be interpreted as the reduced horizon area. 
The entropy has a singularity at  zero reduced horizon area which corresponds to a singular mass which is less than
the remnant mass at which the radiation process stops thereby avoiding the singularity.

%%%%%%%%%%%%%%%%%%%%%%%%%%%%%%%%%%%%
\section*{Acknowledgements} The authors would like to thank the referees for useful comments.

%%%%%%%%%%%%%%%%%%%%%%%%%%%%%%%%%%%%


\begin{thebibliography}{99}
%%%%%%%%%%%%%
\bibitem{str}D.~Amati, M.~Ciafaloni, G.~Veneziano, Phys. Lett. B 216 (1989) 41.
\bibitem{nc}F.~Girelli, E.R.~Livine, D.~Oriti, Nucl. Phys. B 708 (2005) 411; [gr-qc/0406100].
\bibitem{mead}C.A.~Mead, Phys. Rev. D 135 (1964) 849.
\bibitem{hawk1}S.W.~Hawking, Nature (London) 248 (1974) 30.
\bibitem{hawk2}S.W.~Hawking, Commun. Math. Phys. 43 (1975) 199.
\bibitem{bek} J.D.~Bekenstein, Phys. Rev. D 7 (8) (1973) 2333.
\bibitem{adler}R.J.~ Adler, P.~Chen, D.I.~ Santiago, Gen. Rel. Grav. 33 (2001) 2101; [gr-qc/0106080].
\bibitem{barun}B.~Majumder, Phys. Lett. B 703 (2011) 402.
\bibitem{rb}R.~Banerjee, S.~Ghosh, Phys. Lett. B 688 (2010) 224; arXiv:1002.2302 [gr-qc].
\bibitem{das1}S.~Das, E.C.~Vagenas, Phys. Rev. Lett. 101 (2008) 221301; arXiv:0810.5333 [hep-th].
\bibitem{das2}S.~Das, E.C.~Vagenas, Phys. Rev. Lett. 104 (2010) 119002; arXiv:1003.3208 [hep-th].
\bibitem{das3}S.~Das, E.C.~Vagenas, Can. J. Phys. 87 (2009) 233; arXiv:0901.1768 [hep-th].
\bibitem{das4}A.F.~Ali, S.~Das, E.C.~Vagenas, Phys. Lett. B 678 (2009) 497; arXiv:0906.5396 [hep-th].
\bibitem{das5}S.~Das, E.C.~Vagenas, A.F.~Ali, Phys. Lett. B 690 (2010) 407, Erratum-ibid. 692 (2010) 342; 
arXiv:1005.3368 [hep-th].
\bibitem{das6}A.F.~Ali, S.~Das, E.C.~Vagenas,  Phys. Rev. D 84 (2011) 044013; arXiv:1107.3164 [hep-th].
\bibitem{saurya}S.~Das, R.B.~Mann, Phys. Lett. B 704 (2011) 596–599.
\bibitem{das7}S.~Basilakos, S.~Das, E.C.~Vagenas, JCAP 1009 (2010) 027; arXiv:1009.0365 [hep-th].
\bibitem{das8}W.~Chemissany, S.~Das, A.F.~Ali, E.C.~Vagenas, JCAP 1112 (2011) 017; arXiv:1111.7288 [hep-th].
\bibitem{medved}A.J.M.~Medved, E.C.~Vagenas, Phys. Rev. D 70 (2004) 124021.




\end{thebibliography}
 \end{document}